\documentclass[aps,prl,reprint,groupedaddress]{revtex4-1}
\usepackage[pdftex]{hyperref}
\hypersetup{colorlinks,
			linkcolor={blue!75!black!80!yellow},
			citecolor={blue!75!black!80!yellow},
			urlcolor={blue!75!black!80!yellow},
			pdfstartview=FitH}
\usepackage{graphicx}
\usepackage{amsmath}
\usepackage{amssymb}
\usepackage{mathrsfs}
\usepackage{braket}
\usepackage{graphicx}
\usepackage{comment}
\usepackage{xr}
\usepackage{xcolor,soul}
\usepackage{siunitx}
\usepackage[UKenglish]{babel}
\usepackage{txfonts}  
\usepackage{txfontsb} 

\newcommand*\diff{\mathop{}\!\mathrm{d}}

\renewcommand{\Re}{\operatorname{Re}}
\renewcommand{\Im}{\operatorname{Im}}
\newcommand{\iu}{\mathrm{i}}

\newcommand{\e}{\mathrm{e}} 
\newcommand{\rlindex}{\tau}

\begin{document}

\title{Maximal Spontaneous Photon Emission and Energy Loss from Free Electrons}

\author{Yi Yang$^1$}
\email{yiy@mit.edu}
\author{Aviram Massuda$^1$}
\author{Charles Roques-Carmes$^1$}
\author{Steven E. Kooi$^2$}
\author{Thomas Christensen$^1$}
\author{Steven G. Johnson$^1$}
\author{John D. Joannopoulos$^{1,2}$}
\author{Owen D. Miller$^{3}$}
\email{owen.miller@yale.edu}
\author{Ido Kaminer$^{1,4}$}
\email{kaminer@technion.ac.il}
\author{Marin Solja\v{c}i\'{c}$^{1}$}

\affiliation{$^1$Research Laboratory of Electronics, Massachusetts Institute of Technology, Cambridge, Massachusetts 02139, USA}
\affiliation{$^2$Institute for Soldier Nanotechnologies, 77 Massachusetts Avenue, Cambridge, Massachusetts 02139, USA}
\affiliation{$^3$Department of Applied Physics \& Energy Sciences Institute, Yale University, New Haven, CT 06520, USA}
\affiliation{$^4$Andrew \& Erna Viterbi Department of Electrical Engineering, Technion-Israel Institute of Technology, 32000 Haifa, Israel}

\maketitle

\textbf{Free electron radiation such as Cerenkov~\cite{cherenkov34}, Smith--Purcell~\cite{smith1953visible}, and transition radiation~\cite{ginsburg1946radiation,goldsmith1959optical} can be greatly affected by structured optical environments, as has been demonstrated in a variety of polaritonic~\cite{liu2012surface,kaminer2016efficient}, photonic-crystal~\cite{luo2003cerenkov}, and metamaterial~\cite{adamo2009light,ginis2014controlling,liu2017integrated} systems. However, the amount of radiation that can ultimately be extracted from free electrons near an arbitrary material structure has remained elusive. Here we derive a fundamental upper limit to the spontaneous photon emission and energy loss of free electrons, regardless of geometry, which illuminates the effects of material properties and electron velocities. We obtain experimental evidence for our theory with quantitative measurements of Smith--Purcell radiation. Our framework allows us to make two predictions. One is a new regime of radiation operation---at subwavelength separations, slower (nonrelativistic) electrons can achieve stronger radiation than fast (relativistic) electrons. The second is a divergence of the emission probability in the limit of lossless materials. We further reveal that such divergences can be approached by coupling free electrons to photonic bound states in the continuum (BICs)~\cite{hsu2013observation,yang2014analytical,hsu2016bound}. Our findings suggest that compact and efficient free-electron radiation sources from microwaves to the soft X-ray regime may be achievable without requiring ultrahigh accelerating voltages.}

The Smith--Purcell effect epitomizes the potential of free-electron radiation. Consider an electron at velocity $\beta=v/c$ traversing a structure with periodicity $a$; it generates far-field radiation at wavelength $\lambda$ and polar angle $\theta$, dictated by~\cite{smith1953visible}
\begin{equation}
\lambda = \frac{a}{m}\left({\frac{1}{\beta}-\cos\theta}\right),
\label{SP}
\end{equation}
where $m$ is the integer diffraction order. The absence of a minimum velocity in Eq.~\eqref{SP} offers prospects for threshold-free and spectrally tunable light sources,
spanning from microwave and Terahertz~\cite{urata1998superradiant,korbly2005observation,doucas1992first}, across visible~\cite{kube2002observation,yamamoto2015interference,kaminer2017spectrally}, and towards X-ray~\cite{moran1992x} frequencies. In stark contrast to the simple momentum-conservation determination of wavelength and angle, there is no unified yet simple analytical equation for the radiation \textit{intensity}.
Previous theories only offer explicit solutions either under strong assumptions (e.g., assuming perfect conductors or employing effective medium descriptions) or for simple, symmetric geometries ~\cite{van1973smith,haeberle1994calculations,sergeeva2015conical}. Consequently, heavily numerical strategies are often an unavoidable resort~\cite{pendry1994energy,de2000smith}.
The inherent complexity of the interactions between electrons and photonic media have prevented a more general understanding of how pronounced Smith--Purcell radiation and its siblings can ultimately be for \emph{arbitrary} structures, and consequently, how to design the maximum enhancement for free-electron light-emitting devices.

\begin{figure*}[htbp]
\centerline{
\includegraphics[width=0.84\linewidth]{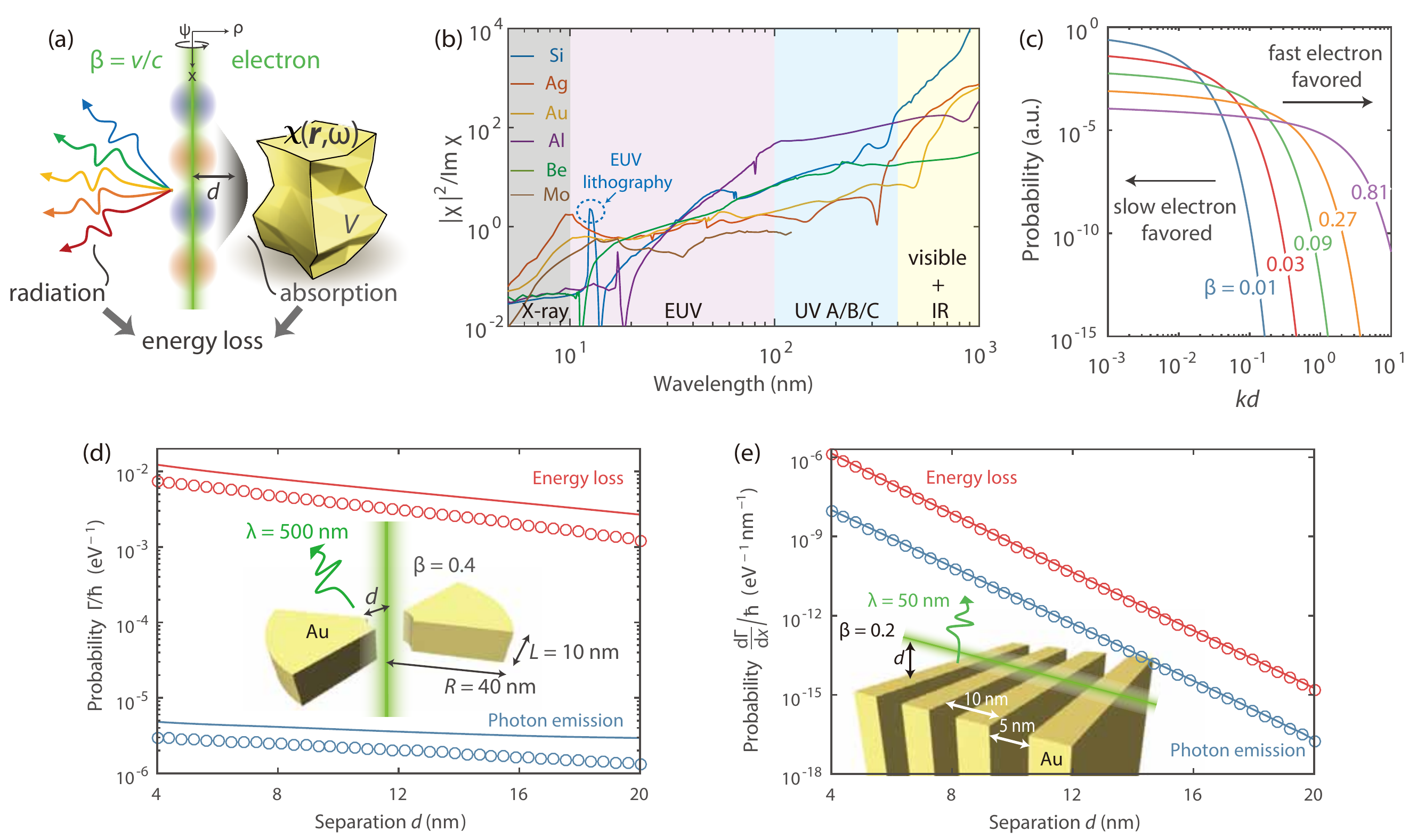}}
 \caption{\textbf{Theoretical framework and predictions.} (a) The interaction between a free electron and  an obstacle defined by a susceptibility tensor $\boldsymbol{\chi}(\mathbf{r},\omega)$ within a volume $V$, located at a distance $d$, generates electron energy loss into radiation and absorption. (b)  $|\chi|^2/\rm Im \chi$ constrains the maximum material response to the optical excitations of free electrons over different spectral ranges for representative materials (from Ref.~\cite{palik1998handbook}) . At the X-ray and EUV regime, Si is optimal near the technologically relevant 13.5~nm (dashed circle). Contrary to the image charge intuition for the optical excitations of electrons, low-loss dielectrics (such as Si in the visible and infrared regimes) can be superior to metals. (c) Shape-independent upper limit showing superiority of slow or fast electrons at small or large separations; the material $\boldsymbol{\chi}$ only affects the overall scaling. (d--e) Numerical simulations (circles) compared to analytical upper limits [lines; Eq.~\eqref{indieBound} for (d) and Eq.~\eqref{SPBound2} for (e), respectively] for the radiation (blue) and energy loss (red) of electrons (d) penetrating the center of an annular bowtie antenna and (e) passing above a grating.}
\label{fig1}
\end{figure*}

We begin our analysis by considering an electron (charge $-e$) of constant velocity
$v\hat{\mathbf{x}}$ traversing a generic scatterer (plasmonic or dielectric, finite or extended) of arbitrary size and material composition, as in Fig.~\ref{fig1}(a). The free current density of the
electron, $\mathbf{J}(\mathbf{r},t) = -\hat{\mathbf{x}}ev\delta(y)\delta(z)\delta(x-vt)$, generates a frequency-dependent ($\e^{-\iu \omega t}$ convention) incident
field~\cite{de2010optical}
\begin{equation}
\mathbf{E}_{\rm inc}(\mathbf{r},\omega)= \frac{e\kappa_{\rho} \e^{\iu k_vx}}{2\pi\omega\epsilon_0}[\hat{\mathbf{x}}\iu \kappa_{\rho} K_0( \kappa_{\rho}\rho) - \hat{\boldsymbol{\rho}} k_vK_1( \kappa_{\rho}\rho)], \label{E3D}
\end{equation}
written in cylindrical coordinates $(x,\rho,\psi)$; here, $K_n$ is the modified Bessel function of the second kind~\cite{abramowitz1964handbook}, $k_v = \omega/v$, and $ \kappa_{\rho} = \sqrt{k_v^2-k^2}=k/\beta\gamma$ ($k=\omega/c$, free-space wavevector; $\gamma=1/\sqrt{1-\beta^2}$, Lorentz factor). Hence, the photon emission and energy loss of free electrons can be treated as a scattering problem: the electromagnetic fields $\mathbf{F_{\rm inc}}=(\mathbf{E_{\rm inc}}, Z_0\mathbf{H_{\rm inc}})^{\rm T}$ (for free-space impedance $Z_0$) are incident upon a photonic medium with material susceptibility $\boldsymbol{\chi}$ (a $6\times6$ tensor for a general medium), causing both absorption and far-field scattering---i.e., photon emission---that together comprise electron energy loss [Fig.~\ref{fig1}(a)].

\begin{figure}[htbp]
	\centering
	\includegraphics[width=0.84\linewidth]{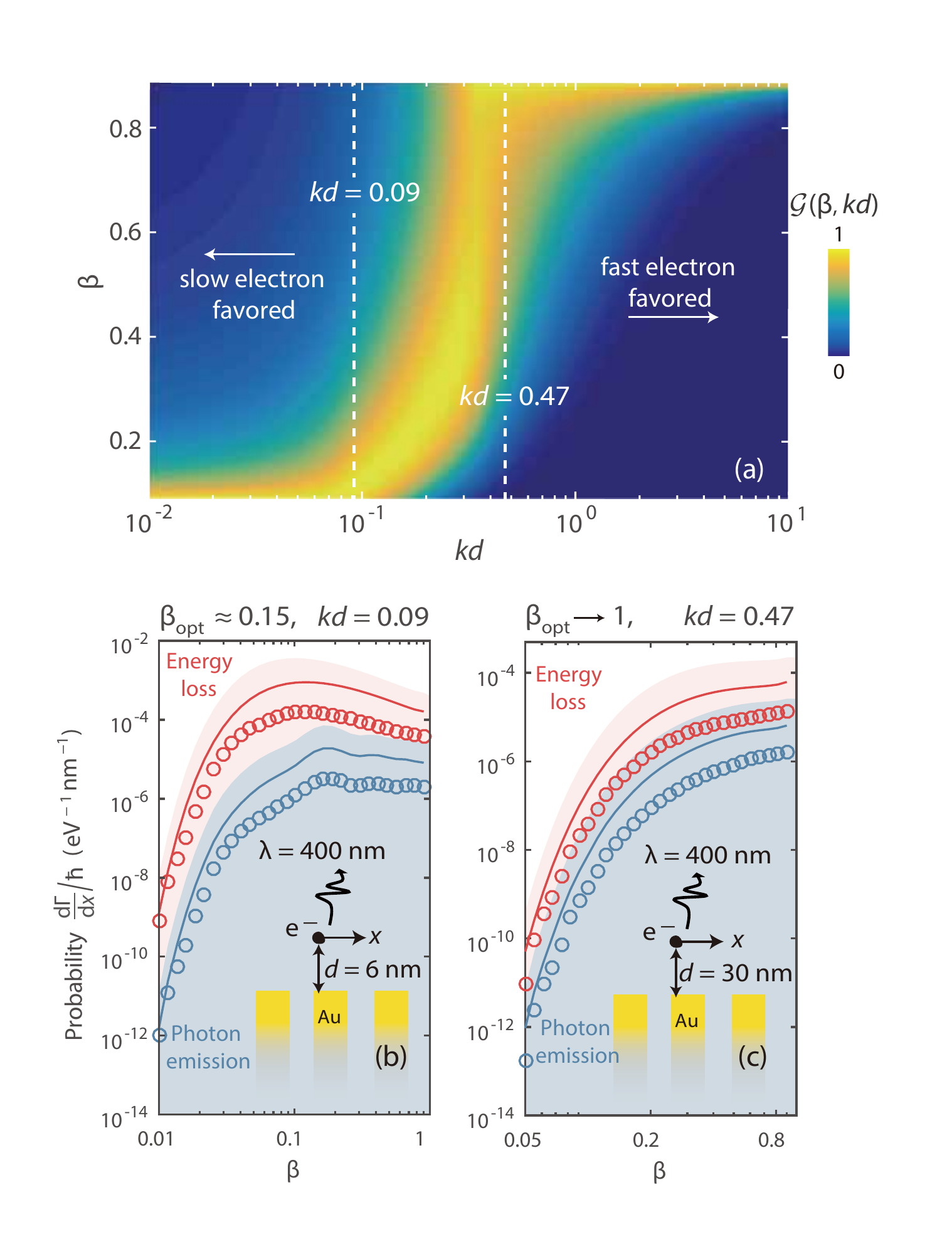}
        \caption{\textbf{Optimal electron velocities for maximal Smith--Purcell radiation.} (a) Behavior of ${\mathcal{G}}\left(\beta,kd\right)$, Eq.~\eqref{SPBound2}, whose maxima indicate separation-dependent optimal electron velocities. Here $\mathcal{G}$ is normalized between 0 and 1 for each separation. The limit yields sharply-contrasting predictions: slow electrons are optimal in the near field ($kd\ll1$) and fast electrons are optimal in the far field ($kd\gg1$). (b--c) Energy loss (red) and radiation (blue) rates [circles: full-wave simulations; lines: grating limit, Eq.~\eqref{SPBound2}; shadings: shape-independent limit, Eq.~\eqref{indieBound_both}] at two representative near/far-field separation distances [white dashed slices in (a)]. }
	\label{fig3}
\end{figure}

As recently shown in Refs.~\cite{miller2016fundamental,yang2017low,miller2015shape}, for a generic electromagnetic scattering problem, passivity---the condition that polarization currents do no net work---constrains the maximum optical response from a given incident field. Consider three power quantities derived from $\mathbf{F}_{\rm inc}$ and the total field $\mathbf{F}$ within the scatterer volume $V$: the total power lost by the electron, $P_{\rm loss}= -(1/2) \Re \int_V \mathbf{J}^*\! \cdot\! \mathbf{E}\diff V =(\epsilon_0\omega/2) \Im \int_V \mathbf{F}_{\rm inc}^{\dag}\boldsymbol{\chi}\mathbf{F}\diff V$, the power absorbed by the medium, $P_{\rm abs}=(\epsilon_0\omega/2) \Im \int_V \mathbf{F}^{\dag}\boldsymbol{\chi}\mathbf{F}\diff V$, and their difference, the power radiated to the far field, $P_{\rm rad} = P_{\rm loss} - P_{\rm abs}$.
Treating $\mathbf{F}$ as an independent variable, the total loss $P_{\rm loss}$ is a \emph{linear} function of $\mathbf{F}$, whereas the fraction that is dissipated is a \emph{quadratic} function of $\mathbf{F}$. Passivity requires nonnegative radiated power, represented by the inequality $P_{\rm abs} < P_{\rm loss}$, which in this framework is therefore a \emph{convex} constraint on any response function. Constrained maximization (see Supplementary 1) of the energy-loss and photon-emission power quantities, $P_{\rm loss}$ and $P_{\rm rad}$, directly yields the limits
\begin{equation}
P_{\tau}(\omega) \leq \frac{\epsilon_0\omega\xi_{\tau}}{2}\int_V \mathbf{F}_{\rm inc}^\dag\boldsymbol{\chi}^{\dag}(\Im \boldsymbol{\chi})^{-1}\boldsymbol{\chi}\mathbf{F}_{\rm inc}\diff V,
\label{scaBound}
\end{equation}
where $\rlindex \in \{\text{rad},\text{loss}\}$ and $\xi_\tau$ accounts for a variable radiative efficiency $\eta$ (defined as the ratio of radiative to total energy loss):
$\xi_{\text{loss}}=1$ and $\xi_{\text{rad}}=\eta(1-\eta)\leq1/4$. Hereafter, we consider isotropic and nonmagnetic materials (and thus a scalar susceptibility $\chi$), but the generalizations to anisotropic and/or magnetic media are straightforward.

Combining Eqs.~\eqref{E3D} and \eqref{scaBound} yields a general limit on the loss or emission spectral probabilities $\Gamma_{\rlindex}(\omega)\!=\!P_{\tau}(\omega)/\hbar\omega$:
\begin{align}
\Gamma_{\rlindex}(\omega) &\leq 
\frac{\alpha\xi_{\rlindex}c}{2\pi\omega^2}\!\!
\int_V \frac{|\chi|^2}{\Im \chi}\left[ \kappa_{\rho}^4K_0^2( \kappa_{\rho}\rho)+ \kappa_{\rho}^2k_v^2K_1^2( \kappa_{\rho}\rho)\right] \diff V,
\label{radBound}
\end{align}
where $\alpha$ is the fine-structure constant.
Equation~\eqref{radBound} imposes, \emph{without} solving Maxwell's equations, a maximum rate of photon generation based on the electron velocity $\beta$ (through $k_v$ and $\kappa_{\rho}$), the material composition $\chi(\mathbf{r})$, and the volume $V$.

The limit in Eq.~\eqref{radBound} can be further simplified by removing the shape dependence of $V$, since the integrand is positive and is thus bounded above by the same integral for any enclosing structure. A scatterer separated from the electron by a minimum distance $d$ can be enclosed within a larger concentric hollow cylinder sector of inner radius $d$ and outer radius $\infty$. For such a sector (height $L$ and opening azimuthal angle $\psi\in[0,2\pi]$), Eq.~\eqref{radBound} can be further simplified, leading to a general closed-form \emph{shape-independent} limit (see Supplementary 2) that highlights the pivotal role of the impact parameter $\kappa_{\rho}d$:
\begin{subequations}
\begin{align}
    \Gamma_{\rlindex}(\omega) &\leq   \frac{\alpha\xi_{\rlindex} }{2\pi c}\frac{|\chi|^2}{\Im \chi} \frac{L\psi}{\beta^2} \left[(\kappa_{\rho} d) K_0( \kappa_{\rho} d)K_1( \kappa_{\rho} d)\right],\label{indieBound}
\\
&\propto \frac{1}{\beta^2}\begin{cases}
    \ln \left(1/\kappa_{\rho} d\right) &\text{for } \kappa_{\rho} d\ll 1, \\
\pi\e^{-2 \kappa_{\rho} d}/2 &\text{for } \kappa_{\rho} d\gg 1.
\end{cases}
\label{indieBound_asymp}
\end{align}
\label{indieBound_both}\end{subequations}
The limits of Eqs.~(\ref{radBound},\ref{indieBound_both}) are completely general; they set the maximum photon emission and energy loss of an electron beam coupled to an arbitrary photonic environment in either the nonretarded or retarded regimes, given only the beam properties and material composition. The key factors that determine maximal radiation are identified: intrinsic material loss (represented by $\Im \chi$), electron velocity $\beta$, and impact parameter $\kappa_{\rho}d$.
The metric $|\chi|^2/\Im\chi$ reflects the influence of the material choice, which depends sensitively on the radiation wavelength [Fig.~\ref{fig1}(b)].
The electron velocity $\beta$ also appears implicitly in the impact parameter $\kappa_{\rho}d = kd/\beta\gamma$,
showing that the relevant length scale is set by the relativistic velocity of the electron.
The impact parameter $\kappa_\rho d$ reflects the influence of the Lorentz contraction $d/\gamma$; a well-known feature of both electron radiation and acceleration~\cite{friedman1988spontaneous,de2010optical,moran1992x}.

A surprising feature of the limits in Eqs~(\ref{radBound},\ref{indieBound_both}) is their prediction for optimal electron velocities. As shown in Fig.~\ref{fig1}(c), when electrons are in the far field of the structure ($\kappa_{\rho} d\gg1$), stronger photon emission and energy loss are achieved by faster electrons---a well-known result. On the contrary, if electrons are in the near field ($\kappa_{\rho} d\ll1$), \emph{slower} electrons are optimal. This contrasting behavior is evident in the asymptotics of Eq.~\eqref{indieBound_asymp}, where the $1/\beta^2$ or $e^{-2\kappa_{\rho}d}$ dependence is dominant at short or large separations. Physically, the optimal velocities are determined by the incident-field properties [Eq.~\eqref{E3D}]: slow electrons generate stronger near field amplitudes although they are more evanescent (Supplementary 2). There has been a recent interest in using low-energy electrons for Cherenkov~\cite{liu2017integrated} and Smith--Purcell~\cite{massuda2017smith} radiation; our prediction that they can be optimal at subwavelength interaction distances underscores the substantial technological potential of nonrelativistic free-electron radiation sources.

The tightness of the limit [Eqs.~(\ref{radBound},\ref{indieBound_both})] is demonstrated by comparison with full-wave numerical calculations (see Methods.) in Figs.~\ref{fig1}(d--e). Two scenarios are considered: in Fig.~{\ref{fig1}(d)}, an electron traverses the center of an annular Au bowtie antenna and undergoes antenna-enabled transition radiation ($\eta\approx0.07\%$), while, in Fig.~{\ref{fig1}(e)}, an electron traverses a Au grating, undergoing Smith--Purcell radiation ($\eta\approx0.9\%$). In both cases, the numerical results closely trail the upper limit at the considered wavelengths, showing that the limits can be approached or even attained with modest effort.

\begin{figure*}[htbp]
	\centering
	\includegraphics[width=0.85\linewidth]{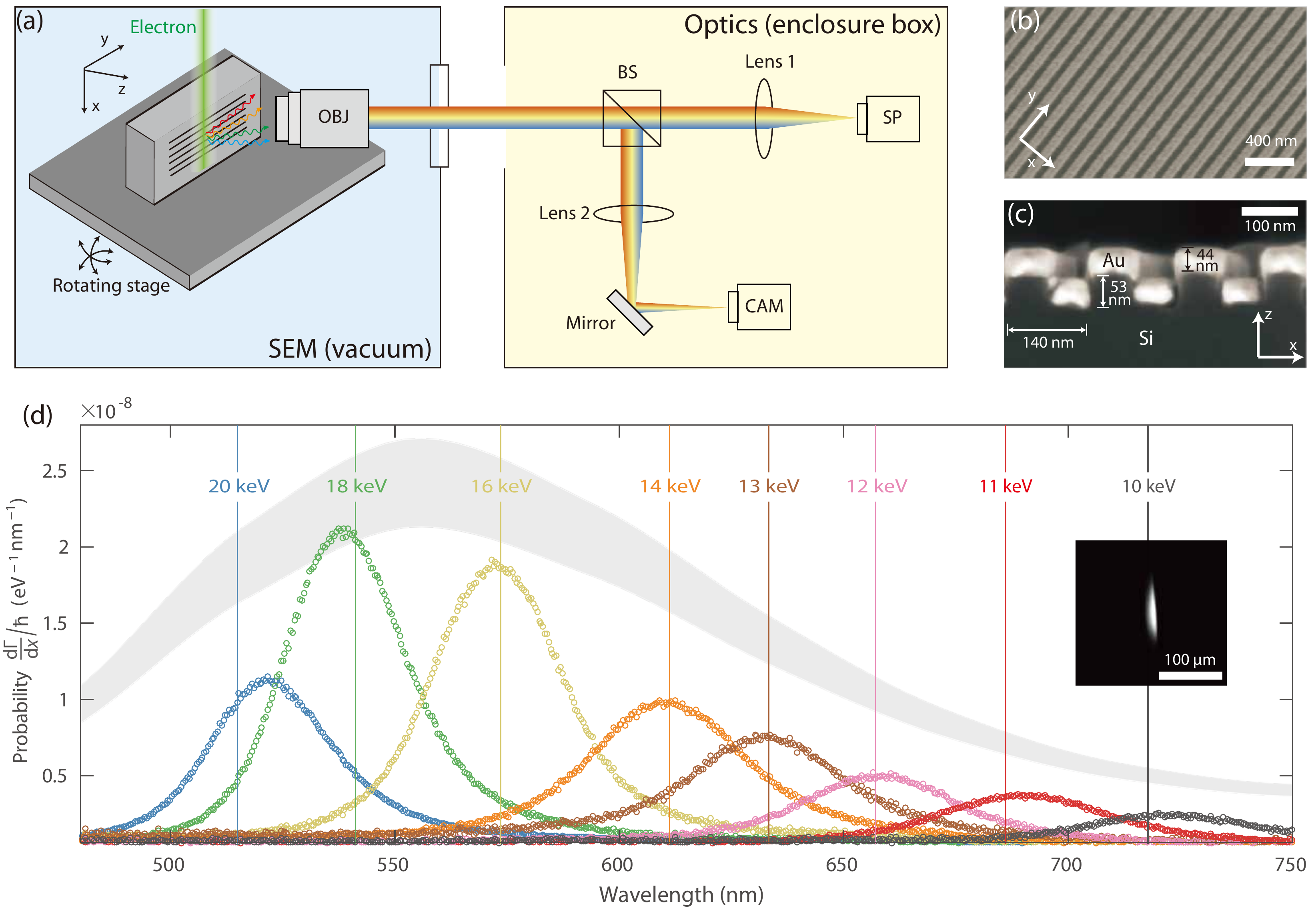}
        \caption{\textbf{Experimental probing of the upper limit.} (a) Experimental setup. OBJ, objective (${\rm NA}=0.3$); BS, beam splitter; SP, spectrometer; CAM, camera. (b-c) SEM images of the structure in (b) top view and (c) cross-sectional view. (d) Quantitative measurement of Smith--Purcell radiation (inset: camera image of the radiation). Solid lines mark the theoretical radiation wavelengths at the normal angle [Eq.~\eqref{SP}]. The envelope (peak outline) of the measured spectra (dots) follows the theoretical upper limit (shaded to account for fabrication tolerance; calculated at each wavelength with the corresponding electron velocity for surface-normal radiation).}
	\label{fig4}
\end{figure*}

Next, we specialize in the canonical Smith--Purcell setup illustrated in Fig.~\ref{fig1}(e) inset. This setup warrants a particularly close study, given its prominent historical and practical role in free-electron radiation. Aside from the shape-independent limit [Eq.~\eqref{indieBound_both}], we can find a sharper limit (in per unit length for periodic structure) specifically for Smith--Purcell radiation using rectangular gratings of filling factor $\Lambda$ (see Supplementary 3)
\begin{equation}
\frac{\diff \Gamma_{\rlindex}(\omega)}{\diff x}
\leq \frac{\alpha\xi_{\rlindex}}{2\pi c}\frac{|\chi|^2}{\Im \chi}\Lambda\mathcal{G}(\beta,kd).
\label{SPBound2}
\end{equation}
The function $\mathcal{G}(\beta,kd)$ is an azimuthal integral (see Supplementary 3) over the Meijer G-function $G_{1,3}^{3,0}$~\cite{abramowitz1964handbook} that arises in the radial integration of the modified Bessel functions $K_{n}$. We emphasize that Eq.~\eqref{SPBound2} is a specific case of Eq.~\eqref{radBound} for grating structures without any approximations and thus can be readily generalized to multi-material scenarios [see Supplementary Eq.~(S37)].

The grating limit [Eq.~\eqref{SPBound2}] exhibits the same asymptotics as Eq.~\eqref{indieBound_both}, thereby reinforcing the optimal-velocity predictions of Fig.~\ref{fig1}(c). The $(\beta,kd)$ dependence of $\mathcal{G}$, see Fig.~\ref{fig3}(a), shows that slow (fast) electrons maximize Smith--Purcell radiation in the small (large) separation regime. We verify the limit predictions by comparison with numerical simulations: At small separations [Figs.~\ref{fig3}(b)], radiation and energy loss peak at velocity $\beta\approx0.15$, consistent with the limit maximum; at large separations [Figs.~\ref{fig3}(c)], both the limit and the numerical results grow monotonically with $\beta$.

The derived upper limit also applies to Cherenkov and transition radiation, as well as bulk loss in electron energy loss spectroscopy (EELS). For these scenarios where electrons enter material bulk, a subtlety arises for the field divergence along the electron's trajectory [$\rho=0$ in Eq.~\eqref{E3D}] within a potentially lossy medium. This divergence, however, can be regularized by introducing natural, system-specific momentum-cutoffs~\cite{de2010optical}, which then directly permits the application of our theory (see Supplementary 6). Meanwhile, there exist additional competing interaction processes (e.g., electrons colliding with individual atoms). However, they typically occur at much smaller length scales.

\begin{figure*}[htbp]
	\centering
	\includegraphics[width=0.85\linewidth]{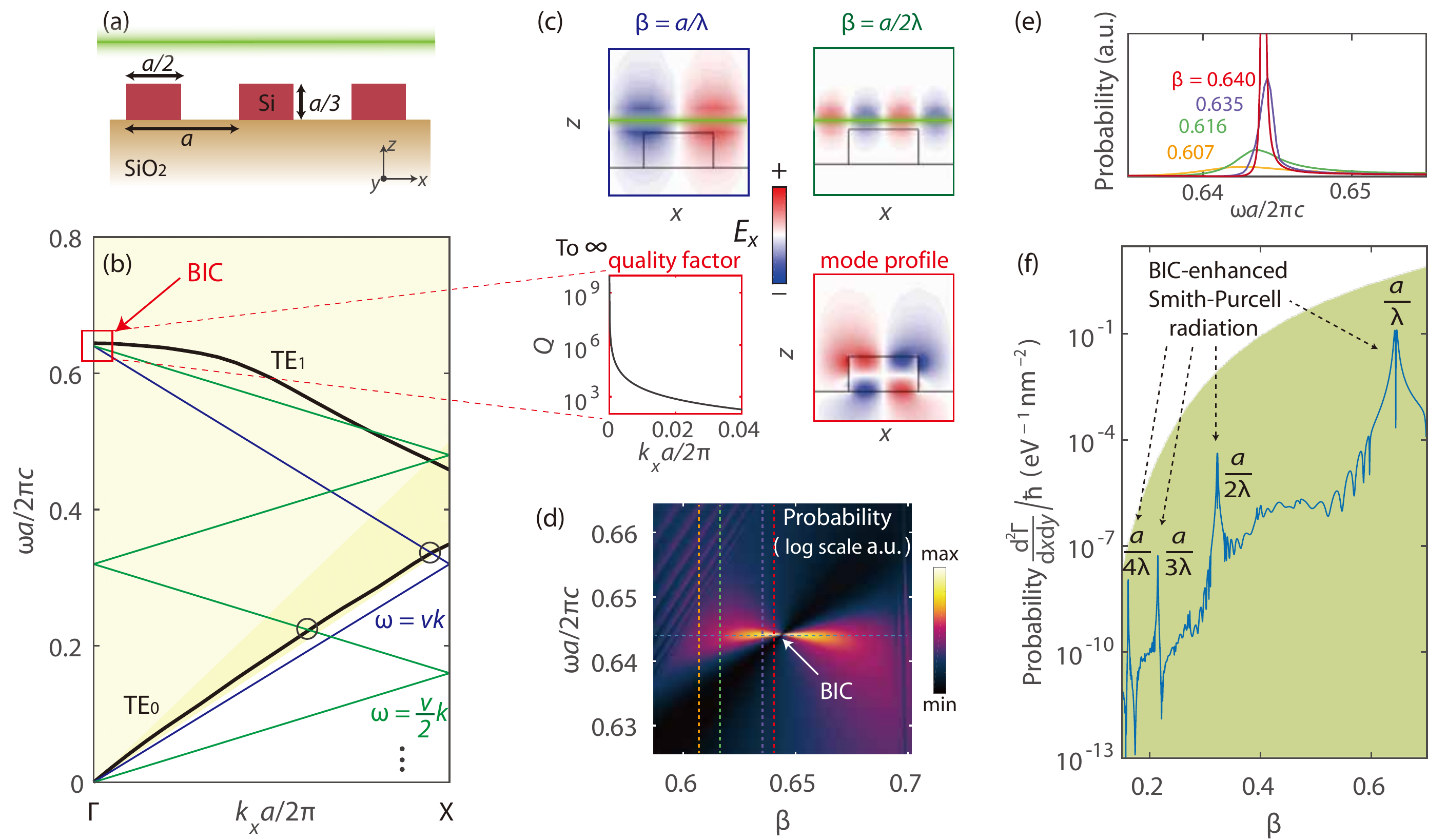}
        \caption{\textbf{Strong enhancement of Smith--Purcell radiation via high-$Q$ resonances near a photonic bound state in the continuum (BIC).} (a) Schematic drawing of a silicon-on-insulator grating (one-dimensional photonic crystal slab: periodic in $x$ and infinite in $y$). (b) Calculated TE band structure (solid black lines) in the $\Gamma$--$X$ direction. The area shaded in light and dark yellow indicates the light cone of air and silica, respectively. The electron lines (blue for velocity $v$, and green for $v/2$) can phase match with either the guided modes (circles) or high-$Q$ resonances near a BIC (red square). (c) Upper: Incident field of electrons. Lower: resonant quality factors (left) and eigenmode profile (right) near a BIC. (d) Strongly enhanced Smith--Purcell radiation near the BIC. (e) Vertical slices of (d). (f) The limit (shaded area) comparing with the horizontal slice of (d), with material loss considered. Strong enhancement happens at electron velocities $\beta=a/m\lambda$ ($m=1,2,3\ldots$).}
	\label{fig5}
\end{figure*}

We perform quantitative experimental measurement of Smith--Purcell radiation to directly probe the upper limit. Fig.~\ref{fig4}(a) shows our experimental setup (see Methods and Supplementary 7 for details).
A one-dimensional 50\%-filling-factor grating (Au-covered single-crystalline Si)---the quintessential Smith--Purcell setup---is chosen as a sample, and shown by SEM images in Figs.~\ref{fig4}(b-c).
Free electrons pass above and impinge onto the sample at a grazing angle of 1.5$^\circ$ under 10 to 20 kV acceleration voltages.

Fig.~\ref{fig4}(d) depicts our measurements of first order $m=1$ Smith--Purcell radiation appearing at wavelengths between 500 and 750 nm. In quantitative agreement with Eq.~\eqref{SP} evaluated at normal emission angle (solid lines), the measured radiation spectra (dots) blueshift with increasing electron velocity. Notably, we experimentally obtain the absolute intensity of the collected radiation via a calibration measurement (see Supplementary 7). The upper limits [Eq.~\eqref{radBound}] for the surface-normal emission wavelengths ($\lambda=a/\beta$) are evaluated at the center of the interaction region [height $\approx140$~nm ($kd\approx1.5$), varying with beam energy], and is shown with shading in Fig.~\ref{fig4}(d) to account for the thickness uncertainty ($\pm1.5$~nm). The envelope spanned by the measurement peaks follows the upper-limit lineshape across the visible spectrum: both the theoretical limit and the measured intensities peak near 550~nm and decrease in a commensurate manner for other wavelengths. This lineshape originates from two competing factors. At shorter wavelengths, the material factor $|\chi^2|/\Im\chi$ decreases significantly for both Au and Si [see Fig.~\ref{fig1}(c)], which accounts for the reduced radiation intensity. At longer wavelengths, the major constraint becomes the less efficient interaction between the electrons and the structure, as the electron-beam diameters increase for the reduced brightness of the electron gun (tungsten) at lower acceleration voltages (see Supplementary~7). These experimental evidences for the upper limit are at $kd\approx1.5$ (estimated from a geometrical ray-tracing model; see Supplementary~7), where fast electrons are still preferred [Fig.~\ref{fig3}(a)]. To further confirm our theory, we also conduct a near-infrared Smith--Purcell experiment (Supplementary~8) at $kd\approx1$, where the envelope lineshape of the emission spectra again follows our prediction. Additionally, we also obtain complementary supporting evidence (extracted from the data in a recent work~\cite{liu2017integrated}) for our slow-electron-efficient prediction (see Supplementary~9).

Finally, we turn our attention to an ostensible peculiarity of the limits: Eq.~\eqref{radBound} evidently diverges for lossless materials ($\Im \chi \rightarrow 0$), seemingly providing little insight. On the contrary, this divergence suggests the existence of a mechanism capable of strongly enhancing Smith--Purcell radiation.
Indeed, by exploiting high-$Q$ resonances near BICs~\cite{hsu2016bound} in photonic crystal slabs, we find that Smith--Purcell radiation can be enhanced by orders of magnitude, when specific frequency, phase, and polarization matching conditions are met.

A one-dimensional silicon ($\chi = 11.25$)-on-insulator (SiO$_2$, $\chi = 1.07$) grating interacting with a sheet electron beam illustrates the core conceptual idea most clearly. The transverse electric (TE) ($E_x$, $H_y$, $E_z$) band structure (lowest two bands labeled TE$_0$ and TE$_1$), matched polarization for a sheet electron beam [Eq.~(S41b)], is depicted in Fig.~\ref{fig5}(b) along the $\Gamma-X$ direction. Folded electron wave vectors, $k_v = \omega/v$, are overlaid for two distinct velocities (blue and green).
Strong electron-photon interactions are possible when the electron and photon dispersions intersect: for instance, $k_v$ and the TE$_0$ band intersect (grey circles) below the air light cone (light yellow shading). However, these intersections are largely impractical: the TE$_0$ band is evanescent in the air region, precluding free-space radiation. Still, analogous ideas, employing similar partially guides modes, e.g., spoof plasmons~\cite{pendry2004mimicking}, have been explored for generating electron-enabled guided waves~\cite{andrews2004gain,kumar2006analysis}.

To overcome this deficiency, we theoretically propose a new mechanism for enhanced Smith--Purcell radiation: coupling of electrons with BICs~\cite{hsu2016bound}. The latter have the extreme quality factors of guided modes but are, crucially, embedded in the radiation continuum, guaranteeing any resulting Smith-Purcell radiation into the far field.
By choosing appropriate velocities $\beta=a/m\lambda$ ($m$ any integer; $\lambda$  the BIC wavelength) such that the electron line (blue or green) intersects the TE$_1$ mode at the BIC [red square in Fig.~\ref{fig5}(b)], the strong enhancements of a guided mode can be achieved in tandem with the radiative coupling of a continuum resonance.
In Fig.~\ref{fig5}(c), the incident fields of electrons and the field profile of the BIC indicate their large modal overlaps. The BIC field profile shows complete confinement without radiation, unlike conventional multipolar radiation modes (see Supplementary Fig.~S9). The $Q$s of the resonances are also provided near a symmetry-protected BIC~\cite{hsu2016bound} at the $\Gamma$ point. Figs.~\ref{fig5}(d) and (e) demonstrate the velocity tunability of BIC-enhanced radiation---as the phase matching approaches the BIC, a divergent radiation rate is achieved.

The BIC-enhancement mechanism is entirely accordant with our upper limits. Practically, silicon has nonzero loss across the visible and near infrared wavelengths. E.g., for a period of $a = 676$ nm, the optimally enhanced radiation wavelength is $\approx$ 1050~nm, at which $\chi_{\rm Si}\approx11.25 + 0.001\rm i$ \cite{green2008self}. For an electron--structure separation of 300~nm, we theoretically show in Fig.~\ref{fig5}(f) the strong radiation enhancements ($>$~3 orders of magnitude) attainable by BIC-enhanced coupling. The upper limit [shaded region; 2D analogue of Eq.~\eqref{radBound}, see Supplementary~10] attains extremely large values due to the minute material loss ($|\chi|^2 / \Im \chi \approx 10^5$); nevertheless, BIC-enhanced coupling enables the radiation intensity to closely approach this limit at several resonant velocities. In the presence of absorptive channel, the maximum enhancement occurs at a small offset from the BIC where the $Q$-matching condition (see Supplementary~11) is satisfied, i.e., equal absorptive and radiative rates of the resonances.

In closing, we have theoretically derived and experimentally probed a universal upper limit to the energy loss and photon emission from free electrons.
The limit depends crucially on the impact parameter $\kappa_{\rho}d$, but \emph{not} on any other detail of the geometry. Hence, our limit applies even to the most complex metamaterials and metasurfaces, given only their constituents.
Surprisingly in the near field slow electrons promise stronger radiation than relativistic ones.
The limit predicts a divergent radiation rate as the material loss rate goes to zero, and we show that BIC resonances enable such staggering enhancements. This is relevant for the generation of coherent Smith--Purcell radiation~\cite{urata1998superradiant,andrews2004gain,kumar2006analysis}. The long lifetime, spectral selectivity, and large field enhancement near a BIC can strongly bunch electrons, allowing them to radiate coherently at the same desired frequency, potentially enabling low-threshold Smith--Purcell free electron lasers. The combination of this mechanism and the optimal velocity prediction reveals prospects of low-voltage yet high-power free-electron radiation sources. In addition, our findings demonstrate a simple guiding principle to maximize the signal-to-noise ratio for EELS through an optimal choice of electron velocity, enabling improved spectral resolution.

The predicted slow-electron-efficient regime still calls for direct experimental validation. We suggest that field-emitter-integrated free-electron devices (e.g.~\cite{liu2017integrated}) are ideal to confirm the prediction due to the achievable small electron-structure separation and high electron beam quality at relatively large currents. Additionally, the microwave or Terahertz frequencies could be suitable testing spectral ranges, where the subwavelength separation requirement is more achievable.

The upper limit demonstrated here is in the spontaneous emission regime for constant-velocity electrons, and can be extended to the stimulated regime by suitable reformulation.
Stronger electron-photon interactions can change electron velocity by a non-negligible amount that alters the radiation. If necessary, this correction can be perturbatively incorporated. In the case of external optical pumping~\cite{schachter1989smith}, the upper limit can be revised by redefining the incident field as the summation of the electron incident field and the external optical field. From a quantum mechanical perspective, this treatment corresponds to stimulated emission from free electrons, which multiplies the limit by the number of photons in that radiation mode.
This treatment could also potentially translate our limit into a fundamental limit for particle acceleration~\cite{peralta2013demonstration,breuer2013laser}, which is the time-reversal of free electron energy loss and which typically incorporates intense laser pumping.
In the multi-electron scenario, the radiation upper limit will be obtained in the case of perfect bunching, where all electrons radiate in phase. In this case, our single-electron limit should be multiplied by the number of electrons to correct for the superradiant nature of such coherent radiation.

\section*{Methods}
Methods, including statements of data availability and any associated
accession codes and references, are available at xxx

\section*{Acknowledgements}
The authors thank fruitful discussions with Prof. Shouzhi Yang, Prof. Chao Peng, Prof. Avi Gover, Prof. Bo Zhen, Dr. Liang Jie Wong, Dr. Xiao Lin, Di Zhu, Tena Dubcek, and Nicholas Rivera. We thank Dr. Paola Rebusco for critical reading and editing of the manuscript. This work was performed in part at the Harvard University Center for Nanoscale Systems (CNS), a member of the National Nanotechnology Coordinated Infrastructure Network (NNCI), which is supported by the National Science Foundation under NSF ECCS award no. 1541959. This work was partly supported by the Army Research Office through the Institute for Soldier Nanotechnologies under contract No. W911NF-18-2-0048 and W911NF-13-D-0001. Y.Y. was partly supported by the MRSEC Program of the National Science Foundation under Grant No. DMR-1419807. T.C. was supported by the Danish Council for Independent Research (Grant No. DFF�C6108-00667). O.D.M. was supported by the Air Force Office of Scientific Research under award number FA9550-17-1-0093. I.K. was partially supported by the Seventh Framework Programme of the European Research Council (FP7- Marie Curie IOF) under grant agreement No. 328853�CMC-BSiCS.
\section*{Author contributions}
Y.Y., O.D.M., I.K., and M.S. conceived the project. Y.Y. developed the analytical models and numerical calculations. A.M. prepared the sample under study. Y.Y., A.M., C.R.-C., S.E.K., and I.K. performed the experiment. Y.Y., T.C., and O.D.M. analyzed the asymptotics and bulk loss of the limit. S.G.J., J.D.J., O.D.M., I.K., and M.S. supervised the project. Y.Y. wrote the manuscript with inputs from all authors.
\section*{Competing interests}
The authors declare no competing interests.
\section*{Additional Information}
\textbf{Supplementary information} is available for this paper at xxx

\textbf{Reprints and permissions information} is available at www.nature.com/reprints.

\textbf{Correspondence and requests for materials} should be addressed to Y.Y., O.D.M., and I.K..

\textbf{Publisher��s note:} Springer Nature remains neutral with regard to jurisdictional claims in
published maps and institutional affiliations.


\begin{thebibliography}{40}%
\makeatletter
\providecommand \@ifxundefined [1]{%
 \@ifx{#1\undefined}
}%
\providecommand \@ifnum [1]{%
 \ifnum #1\expandafter \@firstoftwo
 \else \expandafter \@secondoftwo
 \fi
}%
\providecommand \@ifx [1]{%
 \ifx #1\expandafter \@firstoftwo
 \else \expandafter \@secondoftwo
 \fi
}%
\providecommand \natexlab [1]{#1}%
\providecommand \enquote  [1]{``#1''}%
\providecommand \bibnamefont  [1]{#1}%
\providecommand \bibfnamefont [1]{#1}%
\providecommand \citenamefont [1]{#1}%
\providecommand \href@noop [0]{\@secondoftwo}%
\providecommand \href [0]{\begingroup \@sanitize@url \@href}%
\providecommand \@href[1]{\@@startlink{#1}\@@href}%
\providecommand \@@href[1]{\endgroup#1\@@endlink}%
\providecommand \@sanitize@url [0]{\catcode `\\12\catcode `\$12\catcode
  `\&12\catcode `\#12\catcode `\^12\catcode `\_12\catcode `\%12\relax}%
\providecommand \@@startlink[1]{}%
\providecommand \@@endlink[0]{}%
\providecommand \url  [0]{\begingroup\@sanitize@url \@url }%
\providecommand \@url [1]{\endgroup\@href {#1}{\urlprefix }}%
\providecommand \urlprefix  [0]{URL }%
\providecommand \Eprint [0]{\href }%
\providecommand \doibase [0]{http://dx.doi.org/}%
\providecommand \selectlanguage [0]{\@gobble}%
\providecommand \bibinfo  [0]{\@secondoftwo}%
\providecommand \bibfield  [0]{\@secondoftwo}%
\providecommand \translation [1]{[#1]}%
\providecommand \BibitemOpen [0]{}%
\providecommand \bibitemStop [0]{}%
\providecommand \bibitemNoStop [0]{.\EOS\space}%
\providecommand \EOS [0]{\spacefactor3000\relax}%
\providecommand \BibitemShut  [1]{\csname bibitem#1\endcsname}%
\let\auto@bib@innerbib\@empty
\bibitem [{\citenamefont {Cherenkov}(1934)}]{cherenkov34}%
  \BibitemOpen
  \bibfield  {author} {\bibinfo {author} {\bibfnamefont {P.~A.}\ \bibnamefont
  {Cherenkov}},\ }\href@noop {} {\bibfield  {journal} {\bibinfo  {journal}
  {Dokl. Akad. Nauk SSSR}\ }\textbf {\bibinfo {volume} {2}},\ \bibinfo {pages}
  {451} (\bibinfo {year} {1934})}\BibitemShut {NoStop}%
\bibitem [{\citenamefont {Smith}\ and\ \citenamefont
  {Purcell}(1953)}]{smith1953visible}%
  \BibitemOpen
  \bibfield  {author} {\bibinfo {author} {\bibfnamefont {S.~J.}\ \bibnamefont
  {Smith}}\ and\ \bibinfo {author} {\bibfnamefont {E.}~\bibnamefont
  {Purcell}},\ }\href@noop {} {\bibfield  {journal} {\bibinfo  {journal}
  {Physical Review}\ }\textbf {\bibinfo {volume} {92}},\ \bibinfo {pages}
  {1069} (\bibinfo {year} {1953})}\BibitemShut {NoStop}%
\bibitem [{\citenamefont {Ginsburg}\ and\ \citenamefont
  {Frank}(1946)}]{ginsburg1946radiation}%
  \BibitemOpen
  \bibfield  {author} {\bibinfo {author} {\bibfnamefont {V.}~\bibnamefont
  {Ginsburg}}\ and\ \bibinfo {author} {\bibfnamefont {I.}~\bibnamefont
  {Frank}},\ }\href@noop {} {\bibfield  {journal} {\bibinfo  {journal} {Zh.
  Eksp. Teor. Fiz}\ }\textbf {\bibinfo {volume} {16}},\ \bibinfo {pages} {15}
  (\bibinfo {year} {1946})}\BibitemShut {NoStop}%
\bibitem [{\citenamefont {Goldsmith}\ and\ \citenamefont
  {Jelley}(1959)}]{goldsmith1959optical}%
  \BibitemOpen
  \bibfield  {author} {\bibinfo {author} {\bibfnamefont {P.}~\bibnamefont
  {Goldsmith}}\ and\ \bibinfo {author} {\bibfnamefont {J.}~\bibnamefont
  {Jelley}},\ }\href@noop {} {\bibfield  {journal} {\bibinfo  {journal}
  {Philos. Mag.}\ }\textbf {\bibinfo {volume} {4}},\ \bibinfo {pages} {836}
  (\bibinfo {year} {1959})}\BibitemShut {NoStop}%
\bibitem [{\citenamefont {Liu}\ \emph {et~al.}(2012)\citenamefont {Liu},
  \citenamefont {Zhang}, \citenamefont {Liu}, \citenamefont {Gong},
  \citenamefont {Zhong}, \citenamefont {Zhang},\ and\ \citenamefont
  {Hu}}]{liu2012surface}%
  \BibitemOpen
  \bibfield  {author} {\bibinfo {author} {\bibfnamefont {S.}~\bibnamefont
  {Liu}}, \bibinfo {author} {\bibfnamefont {P.}~\bibnamefont {Zhang}}, \bibinfo
  {author} {\bibfnamefont {W.}~\bibnamefont {Liu}}, \bibinfo {author}
  {\bibfnamefont {S.}~\bibnamefont {Gong}}, \bibinfo {author} {\bibfnamefont
  {R.}~\bibnamefont {Zhong}}, \bibinfo {author} {\bibfnamefont
  {Y.}~\bibnamefont {Zhang}}, \ and\ \bibinfo {author} {\bibfnamefont
  {M.}~\bibnamefont {Hu}},\ }\href@noop {} {\bibfield  {journal} {\bibinfo
  {journal} {Phys. Rev. Lett.}\ }\textbf {\bibinfo {volume} {109}},\ \bibinfo
  {pages} {153902} (\bibinfo {year} {2012})}\BibitemShut {NoStop}%
\bibitem [{\citenamefont {Kaminer}\ \emph {et~al.}(2016)\citenamefont
  {Kaminer}, \citenamefont {Katan}, \citenamefont {Buljan}, \citenamefont
  {Shen}, \citenamefont {Ilic}, \citenamefont {L{\'o}pez}, \citenamefont
  {Wong}, \citenamefont {Joannopoulos},\ and\ \citenamefont
  {Solja{\v{c}}i{\'c}}}]{kaminer2016efficient}%
  \BibitemOpen
  \bibfield  {author} {\bibinfo {author} {\bibfnamefont {I.}~\bibnamefont
  {Kaminer}}, \bibinfo {author} {\bibfnamefont {Y.~T.}\ \bibnamefont {Katan}},
  \bibinfo {author} {\bibfnamefont {H.}~\bibnamefont {Buljan}}, \bibinfo
  {author} {\bibfnamefont {Y.}~\bibnamefont {Shen}}, \bibinfo {author}
  {\bibfnamefont {O.}~\bibnamefont {Ilic}}, \bibinfo {author} {\bibfnamefont
  {J.~J.}\ \bibnamefont {L{\'o}pez}}, \bibinfo {author} {\bibfnamefont {L.~J.}\
  \bibnamefont {Wong}}, \bibinfo {author} {\bibfnamefont {J.~D.}\ \bibnamefont
  {Joannopoulos}}, \ and\ \bibinfo {author} {\bibfnamefont {M.}~\bibnamefont
  {Solja{\v{c}}i{\'c}}},\ }\href@noop {} {\bibfield  {journal} {\bibinfo
  {journal} {Nat. Commun.}\ }\textbf {\bibinfo {volume} {7}} (\bibinfo {year}
  {2016})}\BibitemShut {NoStop}%
\bibitem [{\citenamefont {Luo}\ \emph {et~al.}(2003)\citenamefont {Luo},
  \citenamefont {Ibanescu}, \citenamefont {Johnson},\ and\ \citenamefont
  {Joannopoulos}}]{luo2003cerenkov}%
  \BibitemOpen
  \bibfield  {author} {\bibinfo {author} {\bibfnamefont {C.}~\bibnamefont
  {Luo}}, \bibinfo {author} {\bibfnamefont {M.}~\bibnamefont {Ibanescu}},
  \bibinfo {author} {\bibfnamefont {S.~G.}\ \bibnamefont {Johnson}}, \ and\
  \bibinfo {author} {\bibfnamefont {J.}~\bibnamefont {Joannopoulos}},\
  }\href@noop {} {\bibfield  {journal} {\bibinfo  {journal} {Science}\ }\textbf
  {\bibinfo {volume} {299}},\ \bibinfo {pages} {368} (\bibinfo {year}
  {2003})}\BibitemShut {NoStop}%
\bibitem [{\citenamefont {Adamo}\ \emph {et~al.}(2009)\citenamefont {Adamo},
  \citenamefont {MacDonald}, \citenamefont {Fu}, \citenamefont {Wang},
  \citenamefont {Tsai}, \citenamefont {de~Abajo},\ and\ \citenamefont
  {Zheludev}}]{adamo2009light}%
  \BibitemOpen
  \bibfield  {author} {\bibinfo {author} {\bibfnamefont {G.}~\bibnamefont
  {Adamo}}, \bibinfo {author} {\bibfnamefont {K.~F.}\ \bibnamefont
  {MacDonald}}, \bibinfo {author} {\bibfnamefont {Y.}~\bibnamefont {Fu}},
  \bibinfo {author} {\bibfnamefont {C.}~\bibnamefont {Wang}}, \bibinfo {author}
  {\bibfnamefont {D.}~\bibnamefont {Tsai}}, \bibinfo {author} {\bibfnamefont
  {F.~G.}\ \bibnamefont {de~Abajo}}, \ and\ \bibinfo {author} {\bibfnamefont
  {N.}~\bibnamefont {Zheludev}},\ }\href@noop {} {\bibfield  {journal}
  {\bibinfo  {journal} {Phys. Rev. Lett.}\ }\textbf {\bibinfo {volume} {103}},\
  \bibinfo {pages} {113901} (\bibinfo {year} {2009})}\BibitemShut {NoStop}%
\bibitem [{\citenamefont {Ginis}\ \emph {et~al.}(2014)\citenamefont {Ginis},
  \citenamefont {Danckaert}, \citenamefont {Veretennicoff},\ and\ \citenamefont
  {Tassin}}]{ginis2014controlling}%
  \BibitemOpen
  \bibfield  {author} {\bibinfo {author} {\bibfnamefont {V.}~\bibnamefont
  {Ginis}}, \bibinfo {author} {\bibfnamefont {J.}~\bibnamefont {Danckaert}},
  \bibinfo {author} {\bibfnamefont {I.}~\bibnamefont {Veretennicoff}}, \ and\
  \bibinfo {author} {\bibfnamefont {P.}~\bibnamefont {Tassin}},\ }\href@noop {}
  {\bibfield  {journal} {\bibinfo  {journal} {Phys. Rev. Lett.}\ }\textbf
  {\bibinfo {volume} {113}},\ \bibinfo {pages} {167402} (\bibinfo {year}
  {2014})}\BibitemShut {NoStop}%
\bibitem [{\citenamefont {Liu}\ \emph {et~al.}(2017)\citenamefont {Liu},
  \citenamefont {Xiao}, \citenamefont {Ye}, \citenamefont {Wang}, \citenamefont
  {Cui}, \citenamefont {Feng}, \citenamefont {Zhang},\ and\ \citenamefont
  {Huang}}]{liu2017integrated}%
  \BibitemOpen
  \bibfield  {author} {\bibinfo {author} {\bibfnamefont {F.}~\bibnamefont
  {Liu}}, \bibinfo {author} {\bibfnamefont {L.}~\bibnamefont {Xiao}}, \bibinfo
  {author} {\bibfnamefont {Y.}~\bibnamefont {Ye}}, \bibinfo {author}
  {\bibfnamefont {M.}~\bibnamefont {Wang}}, \bibinfo {author} {\bibfnamefont
  {K.}~\bibnamefont {Cui}}, \bibinfo {author} {\bibfnamefont {X.}~\bibnamefont
  {Feng}}, \bibinfo {author} {\bibfnamefont {W.}~\bibnamefont {Zhang}}, \ and\
  \bibinfo {author} {\bibfnamefont {Y.}~\bibnamefont {Huang}},\ }\href@noop {}
  {\bibfield  {journal} {\bibinfo  {journal} {Nat. Photon.}\ }\textbf {\bibinfo
  {volume} {11}},\ \bibinfo {pages} {289} (\bibinfo {year} {2017})}\BibitemShut
  {NoStop}%
\bibitem [{\citenamefont {Hsu}\ \emph {et~al.}(2013)\citenamefont {Hsu},
  \citenamefont {Zhen}, \citenamefont {Lee}, \citenamefont {Chua},
  \citenamefont {Johnson}, \citenamefont {Joannopoulos},\ and\ \citenamefont
  {Solja{\v{c}}i{\'c}}}]{hsu2013observation}%
  \BibitemOpen
  \bibfield  {author} {\bibinfo {author} {\bibfnamefont {C.~W.}\ \bibnamefont
  {Hsu}}, \bibinfo {author} {\bibfnamefont {B.}~\bibnamefont {Zhen}}, \bibinfo
  {author} {\bibfnamefont {J.}~\bibnamefont {Lee}}, \bibinfo {author}
  {\bibfnamefont {S.-L.}\ \bibnamefont {Chua}}, \bibinfo {author}
  {\bibfnamefont {S.~G.}\ \bibnamefont {Johnson}}, \bibinfo {author}
  {\bibfnamefont {J.~D.}\ \bibnamefont {Joannopoulos}}, \ and\ \bibinfo
  {author} {\bibfnamefont {M.}~\bibnamefont {Solja{\v{c}}i{\'c}}},\ }\href@noop
  {} {\bibfield  {journal} {\bibinfo  {journal} {Nature}\ }\textbf {\bibinfo
  {volume} {499}},\ \bibinfo {pages} {188} (\bibinfo {year}
  {2013})}\BibitemShut {NoStop}%
\bibitem [{\citenamefont {Yang}\ \emph {et~al.}(2014)\citenamefont {Yang},
  \citenamefont {Peng}, \citenamefont {Liang}, \citenamefont {Li},\ and\
  \citenamefont {Noda}}]{yang2014analytical}%
  \BibitemOpen
  \bibfield  {author} {\bibinfo {author} {\bibfnamefont {Y.}~\bibnamefont
  {Yang}}, \bibinfo {author} {\bibfnamefont {C.}~\bibnamefont {Peng}}, \bibinfo
  {author} {\bibfnamefont {Y.}~\bibnamefont {Liang}}, \bibinfo {author}
  {\bibfnamefont {Z.}~\bibnamefont {Li}}, \ and\ \bibinfo {author}
  {\bibfnamefont {S.}~\bibnamefont {Noda}},\ }\href@noop {} {\bibfield
  {journal} {\bibinfo  {journal} {Phys. Rev. Lett.}\ }\textbf {\bibinfo
  {volume} {113}},\ \bibinfo {pages} {037401} (\bibinfo {year}
  {2014})}\BibitemShut {NoStop}%
\bibitem [{\citenamefont {Hsu}\ \emph {et~al.}(2016)\citenamefont {Hsu},
  \citenamefont {Zhen}, \citenamefont {Stone}, \citenamefont {Joannopoulos},\
  and\ \citenamefont {Solja{\v{c}}i{\'c}}}]{hsu2016bound}%
  \BibitemOpen
  \bibfield  {author} {\bibinfo {author} {\bibfnamefont {C.~W.}\ \bibnamefont
  {Hsu}}, \bibinfo {author} {\bibfnamefont {B.}~\bibnamefont {Zhen}}, \bibinfo
  {author} {\bibfnamefont {A.~D.}\ \bibnamefont {Stone}}, \bibinfo {author}
  {\bibfnamefont {J.~D.}\ \bibnamefont {Joannopoulos}}, \ and\ \bibinfo
  {author} {\bibfnamefont {M.}~\bibnamefont {Solja{\v{c}}i{\'c}}},\ }\href@noop
  {} {\bibfield  {journal} {\bibinfo  {journal} {Nat. Rev. Mater.}\ }\textbf
  {\bibinfo {volume} {1}},\ \bibinfo {pages} {16048} (\bibinfo {year}
  {2016})}\BibitemShut {NoStop}%
\bibitem [{\citenamefont {Urata}\ \emph {et~al.}(1998)\citenamefont {Urata},
  \citenamefont {Goldstein}, \citenamefont {Kimmitt}, \citenamefont {Naumov},
  \citenamefont {Platt},\ and\ \citenamefont {Walsh}}]{urata1998superradiant}%
  \BibitemOpen
  \bibfield  {author} {\bibinfo {author} {\bibfnamefont {J.}~\bibnamefont
  {Urata}}, \bibinfo {author} {\bibfnamefont {M.}~\bibnamefont {Goldstein}},
  \bibinfo {author} {\bibfnamefont {M.}~\bibnamefont {Kimmitt}}, \bibinfo
  {author} {\bibfnamefont {A.}~\bibnamefont {Naumov}}, \bibinfo {author}
  {\bibfnamefont {C.}~\bibnamefont {Platt}}, \ and\ \bibinfo {author}
  {\bibfnamefont {J.}~\bibnamefont {Walsh}},\ }\href@noop {} {\bibfield
  {journal} {\bibinfo  {journal} {Phys. Rev. Lett.}\ }\textbf {\bibinfo
  {volume} {80}},\ \bibinfo {pages} {516} (\bibinfo {year} {1998})}\BibitemShut
  {NoStop}%
\bibitem [{\citenamefont {Korbly}\ \emph {et~al.}(2005)\citenamefont {Korbly},
  \citenamefont {Kesar}, \citenamefont {Sirigiri},\ and\ \citenamefont
  {Temkin}}]{korbly2005observation}%
  \BibitemOpen
  \bibfield  {author} {\bibinfo {author} {\bibfnamefont {S.}~\bibnamefont
  {Korbly}}, \bibinfo {author} {\bibfnamefont {A.}~\bibnamefont {Kesar}},
  \bibinfo {author} {\bibfnamefont {J.}~\bibnamefont {Sirigiri}}, \ and\
  \bibinfo {author} {\bibfnamefont {R.}~\bibnamefont {Temkin}},\ }\href@noop {}
  {\bibfield  {journal} {\bibinfo  {journal} {Phys. Rev. Lett.}\ }\textbf
  {\bibinfo {volume} {94}},\ \bibinfo {pages} {054803} (\bibinfo {year}
  {2005})}\BibitemShut {NoStop}%
\bibitem [{\citenamefont {Doucas}\ \emph {et~al.}(1992)\citenamefont {Doucas},
  \citenamefont {Mulvey}, \citenamefont {Omori}, \citenamefont {Walsh},\ and\
  \citenamefont {Kimmitt}}]{doucas1992first}%
  \BibitemOpen
  \bibfield  {author} {\bibinfo {author} {\bibfnamefont {G.}~\bibnamefont
  {Doucas}}, \bibinfo {author} {\bibfnamefont {J.}~\bibnamefont {Mulvey}},
  \bibinfo {author} {\bibfnamefont {M.}~\bibnamefont {Omori}}, \bibinfo
  {author} {\bibfnamefont {J.}~\bibnamefont {Walsh}}, \ and\ \bibinfo {author}
  {\bibfnamefont {M.}~\bibnamefont {Kimmitt}},\ }\href@noop {} {\bibfield
  {journal} {\bibinfo  {journal} {Phys. Rev. Lett.}\ }\textbf {\bibinfo
  {volume} {69}},\ \bibinfo {pages} {1761} (\bibinfo {year}
  {1992})}\BibitemShut {NoStop}%
\bibitem [{\citenamefont {Kube}\ \emph {et~al.}(2002)\citenamefont {Kube},
  \citenamefont {Backe}, \citenamefont {Euteneuer}, \citenamefont {Grendel},
  \citenamefont {Hagenbuck}, \citenamefont {Hartmann}, \citenamefont {Kaiser},
  \citenamefont {Lauth}, \citenamefont {Sch{\"o}pe}, \citenamefont {Wagner},
  \citenamefont {Walcher},\ and\ \citenamefont
  {Kretzschmar}}]{kube2002observation}%
  \BibitemOpen
  \bibfield  {author} {\bibinfo {author} {\bibfnamefont {G.}~\bibnamefont
  {Kube}}, \bibinfo {author} {\bibfnamefont {H.}~\bibnamefont {Backe}},
  \bibinfo {author} {\bibfnamefont {H.}~\bibnamefont {Euteneuer}}, \bibinfo
  {author} {\bibfnamefont {A.}~\bibnamefont {Grendel}}, \bibinfo {author}
  {\bibfnamefont {F.}~\bibnamefont {Hagenbuck}}, \bibinfo {author}
  {\bibfnamefont {H.}~\bibnamefont {Hartmann}}, \bibinfo {author}
  {\bibfnamefont {K.}~\bibnamefont {Kaiser}}, \bibinfo {author} {\bibfnamefont
  {W.}~\bibnamefont {Lauth}}, \bibinfo {author} {\bibfnamefont
  {H.}~\bibnamefont {Sch{\"o}pe}}, \bibinfo {author} {\bibfnamefont
  {G.}~\bibnamefont {Wagner}}, \bibinfo {author} {\bibfnamefont
  {T.}~\bibnamefont {Walcher}}, \ and\ \bibinfo {author} {\bibfnamefont
  {M.}~\bibnamefont {Kretzschmar}},\ }\href@noop {} {\bibfield  {journal}
  {\bibinfo  {journal} {Phys. Rev. E}\ }\textbf {\bibinfo {volume} {65}},\
  \bibinfo {pages} {056501} (\bibinfo {year} {2002})}\BibitemShut {NoStop}%
\bibitem [{\citenamefont {Yamamoto}\ \emph {et~al.}(2015)\citenamefont
  {Yamamoto}, \citenamefont {de~Abajo},\ and\ \citenamefont
  {Myroshnychenko}}]{yamamoto2015interference}%
  \BibitemOpen
  \bibfield  {author} {\bibinfo {author} {\bibfnamefont {N.}~\bibnamefont
  {Yamamoto}}, \bibinfo {author} {\bibfnamefont {F.~J.~G.}\ \bibnamefont
  {de~Abajo}}, \ and\ \bibinfo {author} {\bibfnamefont {V.}~\bibnamefont
  {Myroshnychenko}},\ }\href@noop {} {\bibfield  {journal} {\bibinfo  {journal}
  {Phys. Rev. B}\ }\textbf {\bibinfo {volume} {91}},\ \bibinfo {pages} {125144}
  (\bibinfo {year} {2015})}\BibitemShut {NoStop}%
\bibitem [{\citenamefont {Kaminer}\ \emph {et~al.}(2017)\citenamefont
  {Kaminer}, \citenamefont {Kooi}, \citenamefont {Shiloh}, \citenamefont
  {Zhen}, \citenamefont {Shen}, \citenamefont {L{\'o}pez}, \citenamefont
  {Remez}, \citenamefont {Skirlo}, \citenamefont {Yang}, \citenamefont
  {Joannopoulos},\ and\ \citenamefont
  {Solja{\v{c}}i{\'c}}}]{kaminer2017spectrally}%
  \BibitemOpen
  \bibfield  {author} {\bibinfo {author} {\bibfnamefont {I.}~\bibnamefont
  {Kaminer}}, \bibinfo {author} {\bibfnamefont {S.}~\bibnamefont {Kooi}},
  \bibinfo {author} {\bibfnamefont {R.}~\bibnamefont {Shiloh}}, \bibinfo
  {author} {\bibfnamefont {B.}~\bibnamefont {Zhen}}, \bibinfo {author}
  {\bibfnamefont {Y.}~\bibnamefont {Shen}}, \bibinfo {author} {\bibfnamefont
  {J.}~\bibnamefont {L{\'o}pez}}, \bibinfo {author} {\bibfnamefont
  {R.}~\bibnamefont {Remez}}, \bibinfo {author} {\bibfnamefont
  {S.}~\bibnamefont {Skirlo}}, \bibinfo {author} {\bibfnamefont
  {Y.}~\bibnamefont {Yang}}, \bibinfo {author} {\bibfnamefont {J.}~\bibnamefont
  {Joannopoulos}}, \ and\ \bibinfo {author} {\bibfnamefont {M.}~\bibnamefont
  {Solja{\v{c}}i{\'c}}},\ }\href@noop {} {\bibfield  {journal} {\bibinfo
  {journal} {Phys. Rev. X}\ }\textbf {\bibinfo {volume} {7}},\ \bibinfo {pages}
  {011003} (\bibinfo {year} {2017})}\BibitemShut {NoStop}%
\bibitem [{\citenamefont {Moran}(1992)}]{moran1992x}%
  \BibitemOpen
  \bibfield  {author} {\bibinfo {author} {\bibfnamefont {M.~J.}\ \bibnamefont
  {Moran}},\ }\href@noop {} {\bibfield  {journal} {\bibinfo  {journal} {Phys.
  Rev. Lett.}\ }\textbf {\bibinfo {volume} {69}},\ \bibinfo {pages} {2523}
  (\bibinfo {year} {1992})}\BibitemShut {NoStop}%
\bibitem [{\citenamefont {Van~den Berg}(1973)}]{van1973smith}%
  \BibitemOpen
  \bibfield  {author} {\bibinfo {author} {\bibfnamefont {P.}~\bibnamefont
  {Van~den Berg}},\ }\href@noop {} {\bibfield  {journal} {\bibinfo  {journal}
  {JOSA}\ }\textbf {\bibinfo {volume} {63}},\ \bibinfo {pages} {1588} (\bibinfo
  {year} {1973})}\BibitemShut {NoStop}%
\bibitem [{\citenamefont {Haeberl{\'e}}\ \emph {et~al.}(1994)\citenamefont
  {Haeberl{\'e}}, \citenamefont {Rullhusen}, \citenamefont {Salom{\'e}},\ and\
  \citenamefont {Maene}}]{haeberle1994calculations}%
  \BibitemOpen
  \bibfield  {author} {\bibinfo {author} {\bibfnamefont {O.}~\bibnamefont
  {Haeberl{\'e}}}, \bibinfo {author} {\bibfnamefont {P.}~\bibnamefont
  {Rullhusen}}, \bibinfo {author} {\bibfnamefont {J.-M.}\ \bibnamefont
  {Salom{\'e}}}, \ and\ \bibinfo {author} {\bibfnamefont {N.}~\bibnamefont
  {Maene}},\ }\href@noop {} {\bibfield  {journal} {\bibinfo  {journal} {Phys.
  Rev. E}\ }\textbf {\bibinfo {volume} {49}},\ \bibinfo {pages} {3340}
  (\bibinfo {year} {1994})}\BibitemShut {NoStop}%
\bibitem [{\citenamefont {Sergeeva}\ \emph {et~al.}(2015)\citenamefont
  {Sergeeva}, \citenamefont {Tishchenko},\ and\ \citenamefont
  {Strikhanov}}]{sergeeva2015conical}%
  \BibitemOpen
  \bibfield  {author} {\bibinfo {author} {\bibfnamefont {D.~Y.}\ \bibnamefont
  {Sergeeva}}, \bibinfo {author} {\bibfnamefont {A.}~\bibnamefont
  {Tishchenko}}, \ and\ \bibinfo {author} {\bibfnamefont {M.}~\bibnamefont
  {Strikhanov}},\ }\href@noop {} {\bibfield  {journal} {\bibinfo  {journal}
  {Phys. Rev. ST Accel. Beams}\ }\textbf {\bibinfo {volume} {18}},\ \bibinfo
  {pages} {052801} (\bibinfo {year} {2015})}\BibitemShut {NoStop}%
\bibitem [{\citenamefont {Pendry}\ and\ \citenamefont
  {Martin-Moreno}(1994)}]{pendry1994energy}%
  \BibitemOpen
  \bibfield  {author} {\bibinfo {author} {\bibfnamefont {J.}~\bibnamefont
  {Pendry}}\ and\ \bibinfo {author} {\bibfnamefont {L.}~\bibnamefont
  {Martin-Moreno}},\ }\href@noop {} {\bibfield  {journal} {\bibinfo  {journal}
  {Phys. Rev. B}\ }\textbf {\bibinfo {volume} {50}},\ \bibinfo {pages} {5062}
  (\bibinfo {year} {1994})}\BibitemShut {NoStop}%
\bibitem [{\citenamefont {de~Abajo}(2000)}]{de2000smith}%
  \BibitemOpen
  \bibfield  {author} {\bibinfo {author} {\bibfnamefont {F.~G.}\ \bibnamefont
  {de~Abajo}},\ }\href@noop {} {\bibfield  {journal} {\bibinfo  {journal}
  {Phys. Rev. E}\ }\textbf {\bibinfo {volume} {61}},\ \bibinfo {pages} {5743}
  (\bibinfo {year} {2000})}\BibitemShut {NoStop}%
\bibitem [{\citenamefont {Palik}(1998)}]{palik1998handbook}%
  \BibitemOpen
  \bibfield  {author} {\bibinfo {author} {\bibfnamefont {E.~D.}\ \bibnamefont
  {Palik}},\ }\href@noop {} {\emph {\bibinfo {title} {Handbook of optical
  constants of solids}}},\ Vol.~\bibinfo {volume} {3}\ (\bibinfo  {publisher}
  {Academic press},\ \bibinfo {year} {1998})\BibitemShut {NoStop}%
\bibitem [{\citenamefont {De~Abajo}(2010)}]{de2010optical}%
  \BibitemOpen
  \bibfield  {author} {\bibinfo {author} {\bibfnamefont {F.~G.}\ \bibnamefont
  {De~Abajo}},\ }\href@noop {} {\bibfield  {journal} {\bibinfo  {journal} {Rev.
  Mod. Phys.}\ }\textbf {\bibinfo {volume} {82}},\ \bibinfo {pages} {209}
  (\bibinfo {year} {2010})}\BibitemShut {NoStop}%
\bibitem [{\citenamefont {Abramowitz}\ and\ \citenamefont
  {Stegun}(1964)}]{abramowitz1964handbook}%
  \BibitemOpen
  \bibfield  {author} {\bibinfo {author} {\bibfnamefont {M.}~\bibnamefont
  {Abramowitz}}\ and\ \bibinfo {author} {\bibfnamefont {I.~A.}\ \bibnamefont
  {Stegun}},\ }\href@noop {} {\emph {\bibinfo {title} {Handbook of mathematical
  functions: with formulas, graphs, and mathematical tables}}},\ Vol.~\bibinfo
  {volume} {55}\ (\bibinfo  {publisher} {Courier Corporation},\ \bibinfo {year}
  {1964})\BibitemShut {NoStop}%
\bibitem [{\citenamefont {Miller}\ \emph {et~al.}(2016)\citenamefont {Miller},
  \citenamefont {Polimeridis}, \citenamefont {Reid}, \citenamefont {Hsu},
  \citenamefont {DeLacy}, \citenamefont {Joannopoulos}, \citenamefont
  {Solja{\v{c}}i{\'c}},\ and\ \citenamefont {Johnson}}]{miller2016fundamental}%
  \BibitemOpen
  \bibfield  {author} {\bibinfo {author} {\bibfnamefont {O.~D.}\ \bibnamefont
  {Miller}}, \bibinfo {author} {\bibfnamefont {A.~G.}\ \bibnamefont
  {Polimeridis}}, \bibinfo {author} {\bibfnamefont {M.~H.}\ \bibnamefont
  {Reid}}, \bibinfo {author} {\bibfnamefont {C.~W.}\ \bibnamefont {Hsu}},
  \bibinfo {author} {\bibfnamefont {B.~G.}\ \bibnamefont {DeLacy}}, \bibinfo
  {author} {\bibfnamefont {J.~D.}\ \bibnamefont {Joannopoulos}}, \bibinfo
  {author} {\bibfnamefont {M.}~\bibnamefont {Solja{\v{c}}i{\'c}}}, \ and\
  \bibinfo {author} {\bibfnamefont {S.~G.}\ \bibnamefont {Johnson}},\
  }\href@noop {} {\bibfield  {journal} {\bibinfo  {journal} {Opt. Express}\
  }\textbf {\bibinfo {volume} {24}},\ \bibinfo {pages} {3329} (\bibinfo {year}
  {2016})}\BibitemShut {NoStop}%
\bibitem [{\citenamefont {Yang}\ \emph {et~al.}(2017)\citenamefont {Yang},
  \citenamefont {Miller}, \citenamefont {Christensen}, \citenamefont
  {Joannopoulos},\ and\ \citenamefont {Solja{\v{c}}i{\'c}}}]{yang2017low}%
  \BibitemOpen
  \bibfield  {author} {\bibinfo {author} {\bibfnamefont {Y.}~\bibnamefont
  {Yang}}, \bibinfo {author} {\bibfnamefont {O.~D.}\ \bibnamefont {Miller}},
  \bibinfo {author} {\bibfnamefont {T.}~\bibnamefont {Christensen}}, \bibinfo
  {author} {\bibfnamefont {J.~D.}\ \bibnamefont {Joannopoulos}}, \ and\
  \bibinfo {author} {\bibfnamefont {M.}~\bibnamefont {Solja{\v{c}}i{\'c}}},\
  }\href@noop {} {\bibfield  {journal} {\bibinfo  {journal} {Nano Lett.}\
  }\textbf {\bibinfo {volume} {17}},\ \bibinfo {pages} {3238} (\bibinfo {year}
  {2017})}\BibitemShut {NoStop}%
\bibitem [{\citenamefont {Miller}\ \emph {et~al.}(2015)\citenamefont {Miller},
  \citenamefont {Johnson},\ and\ \citenamefont {Rodriguez}}]{miller2015shape}%
  \BibitemOpen
  \bibfield  {author} {\bibinfo {author} {\bibfnamefont {O.~D.}\ \bibnamefont
  {Miller}}, \bibinfo {author} {\bibfnamefont {S.~G.}\ \bibnamefont {Johnson}},
  \ and\ \bibinfo {author} {\bibfnamefont {A.~W.}\ \bibnamefont {Rodriguez}},\
  }\href@noop {} {\bibfield  {journal} {\bibinfo  {journal} {Phys. Rev. Lett.}\
  }\textbf {\bibinfo {volume} {115}},\ \bibinfo {pages} {204302} (\bibinfo
  {year} {2015})}\BibitemShut {NoStop}%
\bibitem [{\citenamefont {Friedman}\ \emph {et~al.}(1988)\citenamefont
  {Friedman}, \citenamefont {Gover}, \citenamefont {Kurizki}, \citenamefont
  {Ruschin},\ and\ \citenamefont {Yariv}}]{friedman1988spontaneous}%
  \BibitemOpen
  \bibfield  {author} {\bibinfo {author} {\bibfnamefont {A.}~\bibnamefont
  {Friedman}}, \bibinfo {author} {\bibfnamefont {A.}~\bibnamefont {Gover}},
  \bibinfo {author} {\bibfnamefont {G.}~\bibnamefont {Kurizki}}, \bibinfo
  {author} {\bibfnamefont {S.}~\bibnamefont {Ruschin}}, \ and\ \bibinfo
  {author} {\bibfnamefont {A.}~\bibnamefont {Yariv}},\ }\href@noop {}
  {\bibfield  {journal} {\bibinfo  {journal} {Reviews of modern physics}\
  }\textbf {\bibinfo {volume} {60}},\ \bibinfo {pages} {471} (\bibinfo {year}
  {1988})}\BibitemShut {NoStop}%
\bibitem [{\citenamefont {Massuda}\ \emph {et~al.}(2017)\citenamefont
  {Massuda}, \citenamefont {Roques-Carmes}, \citenamefont {Yang}, \citenamefont
  {Kooi}, \citenamefont {Yang}, \citenamefont {Murdia}, \citenamefont
  {Berggren}, \citenamefont {Kaminer},\ and\ \citenamefont
  {Solja{\v{c}}i{\'c}}}]{massuda2017smith}%
  \BibitemOpen
  \bibfield  {author} {\bibinfo {author} {\bibfnamefont {A.}~\bibnamefont
  {Massuda}}, \bibinfo {author} {\bibfnamefont {C.}~\bibnamefont
  {Roques-Carmes}}, \bibinfo {author} {\bibfnamefont {Y.}~\bibnamefont {Yang}},
  \bibinfo {author} {\bibfnamefont {S.~E.}\ \bibnamefont {Kooi}}, \bibinfo
  {author} {\bibfnamefont {Y.}~\bibnamefont {Yang}}, \bibinfo {author}
  {\bibfnamefont {C.}~\bibnamefont {Murdia}}, \bibinfo {author} {\bibfnamefont
  {K.~K.}\ \bibnamefont {Berggren}}, \bibinfo {author} {\bibfnamefont
  {I.}~\bibnamefont {Kaminer}}, \ and\ \bibinfo {author} {\bibfnamefont
  {M.}~\bibnamefont {Solja{\v{c}}i{\'c}}},\ }\href@noop {} {\bibfield
  {journal} {\bibinfo  {journal} {arXiv preprint arXiv:1710.05358}\ } (\bibinfo
  {year} {2017})}\BibitemShut {NoStop}%
\bibitem [{\citenamefont {Pendry}\ \emph {et~al.}(2004)\citenamefont {Pendry},
  \citenamefont {Martin-Moreno},\ and\ \citenamefont
  {Garcia-Vidal}}]{pendry2004mimicking}%
  \BibitemOpen
  \bibfield  {author} {\bibinfo {author} {\bibfnamefont {J.}~\bibnamefont
  {Pendry}}, \bibinfo {author} {\bibfnamefont {L.}~\bibnamefont
  {Martin-Moreno}}, \ and\ \bibinfo {author} {\bibfnamefont {F.}~\bibnamefont
  {Garcia-Vidal}},\ }\href@noop {} {\bibfield  {journal} {\bibinfo  {journal}
  {Science}\ }\textbf {\bibinfo {volume} {305}},\ \bibinfo {pages} {847}
  (\bibinfo {year} {2004})}\BibitemShut {NoStop}%
\bibitem [{\citenamefont {Andrews}\ and\ \citenamefont
  {Brau}(2004)}]{andrews2004gain}%
  \BibitemOpen
  \bibfield  {author} {\bibinfo {author} {\bibfnamefont {H.}~\bibnamefont
  {Andrews}}\ and\ \bibinfo {author} {\bibfnamefont {C.}~\bibnamefont {Brau}},\
  }\href@noop {} {\bibfield  {journal} {\bibinfo  {journal} {Phys. Rev. ST
  Accel. Beams}\ }\textbf {\bibinfo {volume} {7}},\ \bibinfo {pages} {070701}
  (\bibinfo {year} {2004})}\BibitemShut {NoStop}%
\bibitem [{\citenamefont {Kumar}\ and\ \citenamefont
  {Kim}(2006)}]{kumar2006analysis}%
  \BibitemOpen
  \bibfield  {author} {\bibinfo {author} {\bibfnamefont {V.}~\bibnamefont
  {Kumar}}\ and\ \bibinfo {author} {\bibfnamefont {K.-J.}\ \bibnamefont
  {Kim}},\ }\href@noop {} {\bibfield  {journal} {\bibinfo  {journal} {Phys.
  Rev. E}\ }\textbf {\bibinfo {volume} {73}},\ \bibinfo {pages} {026501}
  (\bibinfo {year} {2006})}\BibitemShut {NoStop}%
\bibitem [{\citenamefont {Green}(2008)}]{green2008self}%
  \BibitemOpen
  \bibfield  {author} {\bibinfo {author} {\bibfnamefont {M.~A.}\ \bibnamefont
  {Green}},\ }\href@noop {} {\bibfield  {journal} {\bibinfo  {journal} {Sol.
  Energy Mater. Sol. Cells}\ }\textbf {\bibinfo {volume} {92}},\ \bibinfo
  {pages} {1305} (\bibinfo {year} {2008})}\BibitemShut {NoStop}%
\bibitem [{\citenamefont {Sch{\"a}chter}\ and\ \citenamefont
  {Ron}(1989)}]{schachter1989smith}%
  \BibitemOpen
  \bibfield  {author} {\bibinfo {author} {\bibfnamefont {L.}~\bibnamefont
  {Sch{\"a}chter}}\ and\ \bibinfo {author} {\bibfnamefont {A.}~\bibnamefont
  {Ron}},\ }\href@noop {} {\bibfield  {journal} {\bibinfo  {journal} {Phys.
  Rev. A}\ }\textbf {\bibinfo {volume} {40}},\ \bibinfo {pages} {876} (\bibinfo
  {year} {1989})}\BibitemShut {NoStop}%
\bibitem [{\citenamefont {Peralta}\ \emph {et~al.}(2013)\citenamefont
  {Peralta}, \citenamefont {Soong}, \citenamefont {England}, \citenamefont
  {Colby}, \citenamefont {Wu}, \citenamefont {Montazeri}, \citenamefont
  {McGuinness}, \citenamefont {McNeur}, \citenamefont {Leedle}, \citenamefont
  {Walz}, \citenamefont {Sozer}, \citenamefont {Cowan}, \citenamefont
  {Schwartz}, \citenamefont {Travish},\ and\ \citenamefont
  {Byer}}]{peralta2013demonstration}%
  \BibitemOpen
  \bibfield  {author} {\bibinfo {author} {\bibfnamefont {E.~A.}\ \bibnamefont
  {Peralta}}, \bibinfo {author} {\bibfnamefont {K.}~\bibnamefont {Soong}},
  \bibinfo {author} {\bibfnamefont {R.}~\bibnamefont {England}}, \bibinfo
  {author} {\bibfnamefont {E.~R.}\ \bibnamefont {Colby}}, \bibinfo {author}
  {\bibfnamefont {Z.}~\bibnamefont {Wu}}, \bibinfo {author} {\bibfnamefont
  {B.}~\bibnamefont {Montazeri}}, \bibinfo {author} {\bibfnamefont
  {C.}~\bibnamefont {McGuinness}}, \bibinfo {author} {\bibfnamefont
  {J.}~\bibnamefont {McNeur}}, \bibinfo {author} {\bibfnamefont {K.~J.}\
  \bibnamefont {Leedle}}, \bibinfo {author} {\bibfnamefont {D.}~\bibnamefont
  {Walz}}, \bibinfo {author} {\bibfnamefont {E.~B.}\ \bibnamefont {Sozer}},
  \bibinfo {author} {\bibfnamefont {B.}~\bibnamefont {Cowan}}, \bibinfo
  {author} {\bibfnamefont {B.}~\bibnamefont {Schwartz}}, \bibinfo {author}
  {\bibfnamefont {G.}~\bibnamefont {Travish}}, \ and\ \bibinfo {author}
  {\bibfnamefont {R.~L.}\ \bibnamefont {Byer}},\ }\href@noop {} {\bibfield
  {journal} {\bibinfo  {journal} {Nature}\ }\textbf {\bibinfo {volume} {503}},\
  \bibinfo {pages} {91} (\bibinfo {year} {2013})}\BibitemShut {NoStop}%
\bibitem [{\citenamefont {Breuer}\ and\ \citenamefont
  {Hommelhoff}(2013)}]{breuer2013laser}%
  \BibitemOpen
  \bibfield  {author} {\bibinfo {author} {\bibfnamefont {J.}~\bibnamefont
  {Breuer}}\ and\ \bibinfo {author} {\bibfnamefont {P.}~\bibnamefont
  {Hommelhoff}},\ }\href@noop {} {\bibfield  {journal} {\bibinfo  {journal}
  {Phys. Rev. Lett.}\ }\textbf {\bibinfo {volume} {111}},\ \bibinfo {pages}
  {134803} (\bibinfo {year} {2013})}\BibitemShut {NoStop}%
\end{thebibliography}
\end{document}